\documentclass[10pt,conference]{IEEEtran}

\newif\ifanonymous
\anonymousfalse

\newcommand{\new}[1]{#1}

\usepackage{amsmath}
\usepackage{amssymb}
\usepackage{mathtools}
\usepackage{hyperref}
\usepackage[capitalise, nameinlink]{cleveref}
\usepackage{zebra-goodies}
\usepackage[numbers]{natbib}
\usepackage{tabularx}
\usepackage{booktabs}
\usepackage{multirow}
\usepackage{threeparttable}
\usepackage{flushend} 
\usepackage[switch]{lineno}

\usepackage[nounderscore]{syntax} 
\newenvironment{densegrammar}{%
\begin{small}
\grammarparsep 3pt
\parsep 0pt
\itemsep 0pt
\begin{grammar}
}%
{%
\end{grammar}
\end{small}
}
\def\nonterm#1{\textnormal{$\langle$\emph{#1}$\rangle$}}

\usepackage{tikz}
\usetikzlibrary{automata,positioning,shapes.callouts}

\begin{document}
\title{An analysis of the relationship of input metrics}

\ifanonymous
\author{Anonymous Author(s)}
\else
\author{\IEEEauthorblockN{Addison Crump}
\IEEEauthorblockA{\textit{CISPA Helmholtz Center for Information Security}\\
Saarbr\"ucken, Germany\\
\href{https://orcid.org/0009-0003-3271-3558}{0009-0003-3271-3558}}}
\fi

\renewcommand{\linenumberfont}{\normalfont\tiny\sffamily\color{red}}

\maketitle

\begin{abstract}
\new{%
Input metrics evaluate the progress of testing in terms of features of inputs present in a test suite.
Previous works, as early as the 1950s, established a number of such metrics, but few endeavored to compare them.
This paper does so by utilizing existing methods proposed for other metric classes in partition testing literature.
After defining and reviewing common input metrics, we begin with a short case study revealing that typical empirical comparison strategies are fundamentally insufficient for comparing metrics.
Then, we demonstrate how one rigorously improves a standard metric by defining and implementing $k$-alt-path, a new metric which reduces redundancy while improving sensitivity over $k$-path.
Each of the other common input metrics are then systematically compared before discussing the implications of our findings.
With these contributions, we bring forward partition testing analysis methods that justify and form a strategy for future research in refining input metrics.
}
\end{abstract}

\section{Introduction}

When we quantify the thoroughness of testing, we most often use definitions of input space subdivision formalized by partition testing.
The most common form of subdivision familiar to software testing practitioners is likely code line coverage: which lines of code have been \emph{covered} during the execution of a particular input or set of inputs.
Where code coverage divides the input space based on program features, input metrics divide based on input features.
Such features can only be measured when input features are quantifiable, and these generally form along three categories: \emph{lexical}, by inspecting the sequences of tokens in the input; \emph{syntactic}, by observing structure of the input; and \emph{semantic}, by considering the meaning of the input.

Lexical measures of input coverage were introduced earliest, measuring observed sequences of tokens present in tested inputs.
This is potentially most famously demonstrated by \citet{shannon} for measuring the lexical redundancy of the English language using \emph{$n$-grams}, observations of token sequences of length $n$.
Using the measured $n$-gram distributions for both letters and words, \citeauthor{shannon} demonstrates that one can reconstruct approximations of the English language by sampling over probable sequences of characters or words.

A few years later, \citet{chomsky} publishes his abstractions for languages.
These allow one to define structural rules for languages, encoding \emph{syntax} rather than raw token sequences to capture the higher-level relationships between words.
These grammars eventually become standard for structuring computer-parseable languages due to their alignment with the grammars of human languages~\cite{natlang-cfg} and their efficient parseability.
With such parsers becoming more prevalent, people naturally sought to test them.
The same structures for parsing enabled tools for generation \new{and corresponding metrics for computing test suite completeness}.
\citet{purdom} first defines a generation strategy to exercise grammars by producing \emph{sentences} (i.e., inputs satisfying the grammar) that exercise every production rule of the grammar, \new{implicitly defining rule coverage}.
Later authors explore input generation by randomly exploring grammars~\cite{compiler-sentences}.
Eventually, \citet{laemmel} defines coverage metrics that introduce context sensitivity and define mechanisms for producing sentences that exercise each manifestation of a production rule.
The current standard for syntactic coverage measurement, $k$-path, is a conceptual extension of \citeauthor{laemmel}'s work to extend this sensitivity to arbitrary depths~\cite{kpath} \new{and corresponding generators which optimize for that metric}.

\new{%
Generators for semantically valid inputs are also becoming more prevalent.
The rapid improvement of LLMs has spurred significant interest into LLM-based test generation, including the generation of test data and inputs~\cite{llm-input-survey}.
Other works detail specialized generators for specific languages and targets, most often so for source code generation in compiler and interpreter testing~\cite{csmith,fuzzil}.
More recently, \emph{language-based testing} tools attempt to produce semantically valid inputs for arbitrary input languages by performing constraint solving in the process of input production~\cite{fandango,isla}.
}
\new{While input metrics and generation have historically been inextricably linked}, coverage metrics for semantic qualities of languages are typically target-specific, \new{and few have attempted to define such metrics~\cite{cov-comptest}}.
\new{Instead, these works tend to rely instead on syntactic metrics for input diversity and test suite completeness measurement.}

\new{%
The impetus for this work was a pair of questions: are the syntactic metrics used by modern input generators \emph{optimal}, and are they \emph{sufficient}?
We discovered that these two questions were superseded by one: how does one even compare input metrics?
This paper answers this question with three theoretical contributions:
}
\begin{itemize}
    \item A brief repudiation of previous comparison strategies between metrics,
    \item Extension of existing metric comparison strategies from partition testing literature to input metrics, and
    \item Analysis of how \new{different classes of input metrics align}.
\end{itemize}
\new{Along the way, we optimize the standard $k$-path metric significantly with a new metric, $k$-alt-path, and systematize the relationship between existing input metrics.}

\section{Preliminaries}

Central to this paper are two concepts: test adequacy criteria relationships and context-free grammars.
We begin by reviewing background literature that gives the foundation for \new{comparing input metrics}.
Then, we review some preliminaries in context-free grammars before defining each input metric discussed through the paper.

\subsection{Notation}

In this paper, we will consider sets (distinct unordered lists), multisets (non-distinct unordered lists), and sequences (non-distinct ordered lists), for which notation often overlaps to describe similar, but not identical concepts.
To avoid later confusion, the notation used throughout this paper is defined explicitly here.

A set $\mathcal{S}_1$ is a \emph{subset} of $\mathcal{S}_2$ (denoted $\mathcal{S}_1 \subseteq \mathcal{S}_2$) iff $\mathcal{S}_1$ contains only elements present within $\mathcal{S}_2$, and is a \emph{proper subset} (denoted $\mathcal{S}_1 \subset \mathcal{S}_2$) iff $\mathcal{S}_1 \subseteq \mathcal{S}_2$ and $\mathcal{S}_1 \neq \mathcal{S}_2$.
The set formed by all subsets of a given set $\mathcal{S}$ is denoted $\mathcal{P}(\mathcal{S})$.
Other operators used throughout are consistent with typical set theoretic notation.

A multiset $\mathcal{M}_1$ is a \emph{submultiset} of $\mathcal{M}_2$ iff the number of occurrences of $e \in \mathcal{M}_1$ (denoted $\text{Occs}(\mathcal{M}_1,e)$) is fewer than that of $\mathcal{M}_2$:
\begin{align}
\begin{split}
\mathcal{M}_1 &\subseteq \mathcal{M}_2 \iff \text{Occs}(\mathcal{M}_1,e) \leq \text{Occs}(\mathcal{M}_2,e) \\&\forall e \in \mathcal{M}_1
\end{split}
\end{align}
$\mathcal{M}_1$ is a \emph{proper submultiset} of $\mathcal{M}_2$ (denoted $\mathcal{M}_1 \subset \mathcal{M}_2$) iff $\mathcal{M}_1 \subseteq \mathcal{M}_2$ and $\mathcal{M}_1 \neq \mathcal{M}_2$.
The $\textbf{uniq}$ operator creates the set formed from a given multiset, i.e.:
\begin{gather}
\textbf{uniq}(\mathcal{M}) = \{ e : e \in \mathcal{M} \}
\end{gather}
A multiset $\mathcal{M}_1$ is a \emph{subset} of $\mathcal{M}_2$ iff $\textbf{uniq}(\mathcal{M}_1) \subseteq \textbf{uniq}(\mathcal{M}_2)$.
The number of items present in a multiset $\mathcal{M}$ is denoted $\lvert \mathcal{M} \rvert$.

Finally, a \emph{sequence} $q$ is a \emph{concatenation} of elements, e.g. $q = x_1 \sqcup \dots \sqcup x_n$, where $\sqcup$ is the concatenation operator and each $x$ is some other (potentially empty) subsequence.
When contextually obvious, the concatenation operator is omitted for concision (e.g., $q = x_1x_2 = x_1 \sqcup x_2$).
A sequence $q_1$ is a \emph{subsequence} of $q_2$ (denoted $q_1 \subseteq q_2$) iff there exists some sequences $x_1,x_2$ where $q_2 = x_1q_1x_2$, and is a \emph{proper subsequence} iff $q_1 \subseteq q_2$ and $q_1 \neq q_2$ (denoted $q_1 \subset q_2$).
The length of a sequence $q$ is denoted $\lvert q \rvert$.
The set of all possible sequences formed by members of some set $\mathcal{S}$ is denoted $\mathcal{S}^*$, and all possible sequences of length $n$ is denoted $\mathcal{S}^n$.

\subsection{Partition Testing}\label{sec:partition-testing}

Suppose we want to determine whether a program $X$ satisfies some specification $Y$.
In empirical testing, we can't execute $X$ with the often infinite domain of inputs $I$.
Instead, we select a set of test cases $T \subseteq I$.
Now we ask a second question: is $T$ enough to test $X$ for $Y$?

This is where we use \emph{test adequacy criteria}~\cite{subsumption}.
We say that a test suite $T$ is $C$-adequate if $X$ is ``thoroughly'' tested for $Y$ according to the criterion $C$, denoted $C(T,X,Y)$.
As an example typical to fuzzing: the ``all-edges'' criterion states that $T$ is adequate for testing $X$ for $Y$ if, by executing $T$ on $X$, all (reachable) edges of the control flow graph of the program are covered.
In reality, we rarely actually \emph{achieve} such criteria, but such analysis gives us a foothold to begin comparing criteria.

\new{%
We say that a criterion $C$ is \emph{subdomain-based} if, for each program $X \in \mathcal{X}$ and specification $Y \in \mathcal{Y}$, $C$ component-wise divides the input domain $I$ into a multiset of subdomains $\mathcal{SD}_C(X,Y)$.
To make this notion more concrete, we can consider a function $f_C : I \times \mathcal{X} \times \mathcal{Y} \mapsto \mathcal{P}(F_C(X,Y))$, where $F_C(X,Y)$ are a set of observable \emph{features}, individual elements representing observable behaviors exhibited during execution over $X$, exercising of components of the specification $Y$, or even different aspects of an input $i$ itself.
For example, the ``all-edges'' criterion $\textbf{edges}$ would define $F_\textbf{edges}(X,Y)$ as the set of edges in the program $X$.
$f_\textbf{edges}(i,X,Y)$ then represents, ``what set of edges of $X$ are traversed when executing $i$?''
Each $f \in F_C(X,Y)$ then acts as an index for a single element $D_f \in \mathcal{SD}_C(X,Y)$ such that
}
\begin{equation}
D_f = \{ i \in I : f \in f_C(i,X,Y) \}
\end{equation}

\new{%
$\mathcal{SD}_C(X,Y)$ is often not mutually exclusive nor distinct.
As an example, when a sequence of program regions are only connected by unconditional edges, those edges necessarily correspond to the same set of inputs, but different subdivisions in $\mathcal{SD}_\textbf{edges}(X,Y)$.
}

Partition testing itself is the process of testing $X$ by exercising different \emph{partitions} (more accurately, subdomains) of the input domain such that a representative candidate of each subdomain is present in the test suite $T$.
Since each test $t \in T \subseteq I$ is a member of some submultiset of subdomains, partition testing informs us that a test suite $T$ is $C$-adequate for $X$ and $Y$ if
\begin{align}
\begin{split}
\bigcup \mathcal{SD}_C(X,Y) &= \bigcup S\\ \text{where } S &= \{ D \in \mathcal{SD}_C(X,Y) : D \cap T \neq \emptyset \}
\end{split}
\end{align}

\subsubsection{Criterion relationships}

A test criterion $C_1$ \emph{subsumes} $C_2$ for $X,Y$ if
\begin{align}
&C_1(T, X, Y) \implies C_2(T, X, Y) \quad \forall T
\end{align}

$C_1$ \emph{universally} subsumes $C_2$ if this applies for all $X$ and $Y$.
In other words: the subsumption relationship tells us when one adequacy criterion supersedes another; satisfying the former necessarily means that you have satisfied the latter.
This relationship is unfortunately quite weak; it only informs us as to the relationship of $C_1$ and $C_2$ in the \emph{extreme} cases of test adequacy, when all subdomains are satisfied.

To allow us to inspect the relationship between \new{two metrics}, we also consider the \emph{covers} relationship~\cite{subsumption}.
We say that $C_1$ covers $C_2$ for $X,Y$ if:
\begin{align}
\begin{split}
&\forall D \in \mathcal{SD}_{C_2}(X,Y) \quad \exists \mathcal{M} \subseteq \mathcal{SD}_{C_1}(X,Y) \\
&\quad \bigcup_{D' \in \mathcal{M}} D' = D
\end{split}
\end{align}
In other words, $C_1$ covers $C_2$ if $C_2$'s subdomains are each covered by some submultiset of subdomains of $C_1$.
As a result, if $C_1$ covers $C_2$, then it necessarily subsumes as well.
Each of these relationships are necessarily transitive and reflexive.
Later, we show relationships across a variety of input metrics.

\subsection{Context-Free Grammars}

A context-free grammar (CFG) $G$ is a 4-tuple $(V,\Sigma,R,S)$.
$V$ and $\Sigma$ are sets of symbols such that $V \cap \Sigma = \emptyset$ representing \emph{nonterminals} and \emph{terminals}, respectively.
$S \in V$ is the ``start symbol'', which we will revisit shortly.
$R \subseteq V \times (V \cup \Sigma)^*$ is a relation.
For some $v \in V$ and $x \in (V \cup \Sigma)^*$, we say that $v$ ``directly expands'' to $x$ iff $(v,x) \in R$ (denoted $v \rightarrow_R x$).
Put simply: $R$ defines the ``rules'' under which each nonterminal (here, $v$) may be expanded into some concatenation of other symbols (here, $x$).

We can broaden the notion of direct expansion to that of \emph{indirect} expansion.
Given sequences $x = x_1 \sqcup \dots \sqcup x_n$ and $y = y_1 \sqcup \dots \sqcup y_n$ (where $\forall k \in [1,n]$, $x_k \in (V \cup \Sigma)$ and $y_k \in (V \cup \Sigma)^*$), we say that $x$ \emph{indirectly} expands to $y$, written $x \Rightarrow_R y$, iff $x_k \rightarrow_R y_k$ when $x_k \in V$ or $x_k = y_k$, otherwise.

We may now apply this indirect definition of expansion recursively; for $x,z \in (V \cup \Sigma)^*$, $x \Rightarrow_R^* z$ ($x$ \emph{recursively expands to} $z$ by $R$) if there exists some \emph{derivation} sequence $x = a_1 \Rightarrow_R  \dots  \Rightarrow_R a_n = z$.
Coming all the way back around now: we say that some input $i \in \Sigma^*$ is in the grammar $G$ iff $S \Rightarrow_R^* i$.
$G$ is \emph{unambiguous} if there exists exactly one or zero sequences of expansions by which this holds for every $i$.
Unless explicitly mentioned otherwise, we only consider such unambiguous grammars for the remainder of the paper.

\subsubsection{Rule syntax and derivation trees}

\begin{figure}
    \centering
    \begin{densegrammar}
        <expr> ::= <number> `+' <expr> | <number>
        
        <number> ::= `0' | <non_zero><digits>
        
        <digits> ::= <digit><digits> | `'
                     
        <digit> ::= `0' | <non_zero>

        <non\_zero> ::= `1' | `2' | `3' | `4' | `5' | `6' | `7' | `8' | `9'
    \end{densegrammar}
    \caption{The ``addition'' grammar, where $S = \nonterm{expr}$.}
    \label{fig:addition-grammar}
\end{figure}

To define grammars, all we need to define are the rules.
Consider \cref{fig:addition-grammar}; here, quoted values are terminals, $\nonterm{v} ::= x$ means $(v,x) \in R$, and $\nonterm{v} ::= x_1 |  \dots  | x_n$ (an ``alternation'' over $x_1, \dots ,x_n$) means that $\forall k \in [1,n]$, $(v,x_k) \in R$, where $x_k$ is the ``$k$-th variant'' of $v$.
Many flavors of context-free grammar definition languages include extensions like nested alternation, where one may define sub-variants of a particular rule (e.g., $x_1 | (x_2 | x_3)$); nested concatenation, used in conjunction with nested alternations to define rules with common subsequences (e.g., $x_1 \sqcup (x_2 \sqcup x_3 | x_4)$); and quantifiers, which allow for convenient repetition of symbols in rules.
All of these primitives can be converted back into the typical rule definition described above by the creation of intermediary nonterminals, though we do not explicitly perform this in this paper.

\begin{figure}
    \centering
    \small
    \begin{tikzpicture}[on grid,auto,node distance=1.5cm]
        \node[state] (expr) {$\nonterm{expr}_1$};
        \node[state] (plus) [below=2cm of expr] {``$+$''};
        \node[state,dashed] (plus-4) [below=4cm of plus] {``$+$''};
        \node[state] (expr-2) [right=of plus] {$\nonterm{expr}_2$};
        \node[state] (number-2) [below=2cm of expr-2] {$\nonterm{number}_2$};
        \node[state] (number-zero-2) [below=2cm of number-2] {``$0$''};

        \node[state] (number) [left=3.375cm of plus] {$\nonterm{number}_1$};
        \node[state] (digits) [left=3.75cm of number-2] {$\nonterm{digits}_1$};
        \node[state] (non-zero) [left=2.25cm of digits] {$\nonterm{non\_zero}_1$};
        \node[state] (five) [below=2cm of non-zero] {``$5$''};
        \node[state] (epsilon) [below=2cm of digits] {``''};

        \coordinate[below=1cm of expr] (line-intermediary);
        \coordinate[left=6.25cm of line-intermediary] (l1);
        \coordinate[right=2.6cm of line-intermediary] (r1);
        \coordinate[below=2cm of l1] (l2);
        \coordinate[below=2cm of r1] (r2);
        \coordinate[below=2cm of l2] (l3);
        \coordinate[below=2cm of r2] (r3);

        \node (a1) [above=1cm of l1,anchor=west] {$S = a_1$};
        \node (a2) [below=1cm of l1,anchor=west] {$a_2$};
        \node (a3) [below=1cm of l2,anchor=west] {$a_3$};
        \node (a4) [below=1cm of l3,anchor=west] {$i = a_4$};

        \path[-,dotted]
            (l1) edge (r1)
            (l2) edge (r2)
            (l3) edge (r3)
            ;

        \path[-,dashed]
            (plus) edge (plus-4)
            ;

        \path[->]
            (expr) edge (number)
            (expr-2) edge (number-2)
            (expr) edge (expr-2)
            (expr) edge (plus)
            (expr-2) edge (number-2)
            (number-2) edge (number-zero-2)
            (number) edge (non-zero)
            (number) edge (digits)
            (non-zero) edge (five)
            (digits) edge (epsilon)
            ;
    \end{tikzpicture}
    \caption{Derivation tree of the input $5+0$. Dashed components are added for clarity on expansion but are not part of the derivation tree.}
    \label{fig:zero-five-tree}
\end{figure}
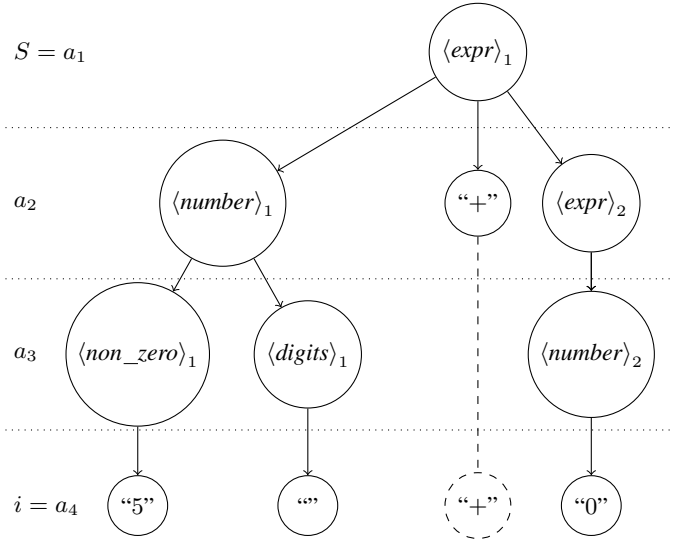

\cref{fig:addition-grammar} defines a grammar that accepts any string that represents the addition of one-to-many nonnegative integers.
To better visualize what this means for inputs, we introduce the concept of a \emph{derivation tree}.
A derivation tree is a representation of the sequence taken during the recursive expansion of the start nonterminal to the corresponding input.
Consider an input $i \in \Sigma^*$ with the value $5+0$.
To show $S \Rightarrow_R^* i$, we present the derivation tree in \cref{fig:zero-five-tree}.
This figure is annotated to show the expansion at each step of $S = a_1 \Rightarrow_R \dots \Rightarrow_R a_4 = i$.
For example, at a tree depth of 2, we see the expansion $a_2 = \nonterm{number}_1 \sqcup \text{``+''} \sqcup \nonterm{expr}_2$.
The derivation tree bridges the formal definition and intuitive understanding of how expansion of a nonterminal actually works in practice.

\subsubsection{Graph representation} \label{sec:graph-construction}

\begin{figure}
    \centering
    \begin{tikzpicture}[on grid,auto,node distance=1.5cm]
        \node[state] (expr) {$\nonterm{expr}_1$};
        \node[state] (expr-alt) [below=of expr] {$|_1$};
        \node[state] (expr-concat-0) [right=2cm of expr-alt] {$\sqcup_1$};
        \node[state] (expr-1) [above=of expr-concat-0] {$\nonterm{expr}_2$};
        \node[state] (number) [below=1.75cm of expr-alt] {$\nonterm{number}_1$};
        \node[state] (number-1) [below=1.75cm of expr-concat-0] {$\nonterm{number}_2$};
        \node[state] (plus) [right=of expr-concat-0] {``$+$''};
        \node[state] (number-alt) [below=1.75cm of number] {$|_2$};
        \node[state] (number-zero) [right=of number-alt] {``$0$''$\text{}_1$};
        \node[state] (number-concat-1) [below=of number-alt] {$\sqcup_2$};
        \node[state] (digits) [right=of number-concat-1] {$\nonterm{digits}_1$};
        \node[state] (digits-alt) [below=of digits] {$|_3$};
        \node[state] (digits-epsilon) [below=of digits-alt] {``''};
        \node[state] (digits-concat) [right=2cm of digits-alt] {$\sqcup_3$};
        \node[state] (digits-1) [above=of digits-concat] {$\nonterm{digits}_2$};
        \node[state] (digit) [below=of digits-concat] {$\nonterm{digit}$};
        \node[state] (digit-alt) [below=of digit] {$|_4$};
        \node[state] (number-zero-2) [right=of digit-alt] {``$0$''$\text{}_2$};
        \node[state] (nonzero) [left=1.75cm of number-concat-1] {$\nonterm{nonzero}_1$};
        \node[state] (nonzero-1) [left=1.75cm of digit-alt] {$\nonterm{nonzero}_2$};
        \node[state] (nonzero-alt) [below=3cm of nonzero] {$|_5$};
        \node[state] (nonzero-1-lit) [below=of nonzero-alt] {``$1$''};
        \node[state,dotted] (nonzero-dots) [right=of nonzero-1-lit] {``$\dots{}$''};
        
        \path[->]
            (expr) edge (expr-alt)
            (expr-alt) edge (expr-concat-0)
            (expr-alt) edge (number)
            (expr-concat-0) edge (number-1)
            (expr-concat-0) edge (plus)
            (expr-concat-0) edge (expr-1)
            (expr-1) edge (expr-alt)
            (number) edge (number-alt)
            (number-1) edge (number-alt)
            (number-alt) edge (number-zero)
            (number-alt) edge (number-concat-1)
            (number-concat-1) edge (nonzero)
            (number-concat-1) edge (digits)
            (digits) edge (digits-alt)
            (digits-alt) edge (digits-epsilon)
            (nonzero) edge (nonzero-alt)
            (digits-alt) edge (digits-concat)
            (digits-concat) edge (digit)
            (digits-concat) edge (digits-1)
            (digits-1) edge (digits-alt)
            (digit) edge (digit-alt)
            (digit-alt) edge (number-zero-2)
            (digit-alt) edge (nonzero-1)
            (nonzero-alt) edge (nonzero-1-lit)
            (nonzero-1) edge[out=135,in=0] (nonzero-alt)
            ;
        
        \path[->,dotted] 
            (nonzero-alt) edge (nonzero-dots)
            ;
    \end{tikzpicture}
    \caption{The graph constructed for the ``addition'' grammar.}
    \label{fig:addition-graph}
\end{figure}
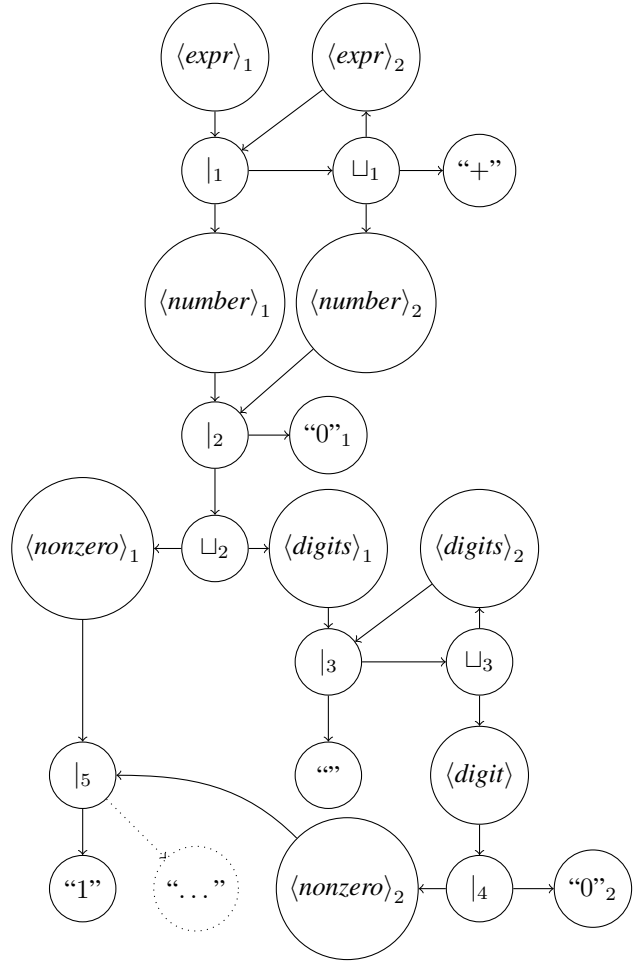

To better visualize the CFG itself, we may represent it as a directed graph.
The original $k$-path paper~\cite{kpath} does so in the following steps\footnote{\citet{kpath} additionally provide steps to represent \emph{quantifiers} (an extension to simplify repetition) within grammar graphs, but do not specify how to compute the $k$-path metric in the presence of quantifiers. To avoid underspecification, we only consider grammars that do not contain quantifiers.}:
\begin{enumerate}
    \item Recursing from the starting nonterminal, the immediate expansion of each nonterminal is added to the graph as children of that nonterminal with the remaining steps.
    \item At each alternation, a unique synthetic $|$ node is inserted and an edge added from this node to each of its variants.
    This step allows for nested alternations to be represented without intermediary nonterminals.
    \item At each concatenation, a unique synthetic $\sqcup$ node is inserted and an edge added from this node to each of the concatenated values.
    This step allows for nested concatenation to be represented without intermediary nonterminals.
    \item Literals are inserted as nodes uniquely.
    \item Each reference to a nonterminal is assigned a unique node, then an edge added to its previous expansion.
\end{enumerate}
A graph for the grammar of \cref{fig:addition-grammar} is shown in \cref{fig:addition-graph}.


\subsection{Traditional Input Metrics}\label{sec:metric-def}

Let $I \subseteq \Sigma^*$ be the set of inputs expressed by an unambiguous context-free grammar $G = (V, \Sigma, R, S)$:
\begin{align}
I = \{ i : S \Rightarrow_R^* i \}
\end{align}
We define coverage as some function $\textbf{cov} : I \mapsto \mathcal{P}(F_C)$, \new{where $F_C$ is the coverage-specific feature domain described in \cref{sec:partition-testing}, absent a specific program $X$ and specification $Y$ as $F_C$ is independent of these for input metrics}.

\subsubsection{$n$-gram Coverage}
The first coverage metric introduced~\cite{shannon}, though not formally as a metric, is a lexical coverage.
$n$-gram coverage ($\textbf{gram}_n : I \mapsto \mathcal{P}(\Sigma^n)$) is defined as the set of sequences of terminals of length $n$ present in a given $i$:
\begin{align}
\textbf{gram}_n(i) = \{ s \in \Sigma^n : s \subseteq i \}
\end{align}
For the purposes of comparison later, we define a second metric, $n$-or-less-gram ($\textbf{gram}_{\leq n} : I \mapsto \mathcal{P}(\Sigma^n \cup \Sigma^{n-1} \cup \dots \cup \Sigma)$), that is inclusive of shorter sequences:
\begin{align}
\textbf{gram}_{\leq n}(i) = \{ s \in \Sigma^* : \lvert s \rvert \leq n \land s \subseteq i \}
\end{align}

\subsubsection{Rule Coverage}
\citet{purdom} defines production rule coverage similarly straightforwardly, intuitively representing the covered expansions of each nonterminal:
\begin{gather}
\begin{split}
\textbf{rule}(i) = &\{ (v,x) \in R \\
&\quad : (\exists u,w \in (V \cup \Sigma)^*) \\
&\quad\quad\quad [S \Rightarrow_R^* uvw \Rightarrow_R uxw \Rightarrow_R^* i] \}
\end{split}
\end{gather}

\subsubsection{Context-Dependent Rule Coverage (CDRC)}
CDRC as defined by \citet{laemmel} is effectively a chained version of \textbf{rule}, wherein we track which \emph{sequences} of expansions are observed.
This allows one to not only measure which production rules are covered, but the context (i.e. the ``parent'' production rule) in which those production rules are used.
$\textbf{cdrc}_2$ is defined accordingly~\cite{laemmel}:
\begin{gather}
\begin{split}
\textbf{cdrc}_2(i) = \, &\{ (v_1,u_1v_2w_1) \sqcup (v_2,x) \in R^2 \\
&\quad : (\exists u,w \in (V \cup \Sigma)^*) \\
&\quad\quad\quad [S \Rightarrow_R^* uv_1w \Rightarrow_R uu_1v_2w_1w \\
&\quad\quad\quad\quad \Rightarrow_R uu_1xw_1w \Rightarrow_R^* i] \}
\end{split}
\end{gather}

For purposes of comparison later, we generalize this coverage to chains of arbitrary length (for $n > 2$):
\begin{align}
\textbf{cdrc}_n(i) = \, &\{ (v_1,u_1v_2w_1) \sqcup \dots \sqcup (v_n,x) \in R^n \notag \\
&\quad : (\exists u,w \in (V \cup \Sigma)^*) \\
&\quad\quad [S \Rightarrow_R^* uv_1w \Rightarrow_R uu_1v_2w_1w \Rightarrow_R \dots \notag \\
&\quad\quad\quad \Rightarrow_R uu_1u_{\dots}u_{n-1}xw_{n-1}w_{\dots}w_1w \Rightarrow_R^* i] \} \notag \\
&\cup \textbf{cdrc}_{n-1}(i) \notag
\end{align}
Note especially the union with $\textbf{cdrc}_{n-1}$; without this, $\textbf{cdrc}_n$ would fail to capture coverage over expansion sequences shorter than $n$.
Note also that neither rule coverage nor CDRC have the ability to represent nested alternations or concatenations.

\subsubsection{$k$-path Coverage}

For the grammar graph $H = (N, E)$ with nodes $N$ and edges $E \subseteq N \times N$, the paths of this graph $P \subseteq N^+$ are each sequences of nodes where:
\begin{align} \label{eq:paths}
\begin{split}
P = &\, \{ x_1 \sqcup \dots \sqcup x_n \in N^+ \\
    &\quad : (\forall k \in [1, n))[(x_k, x_{k+1}) \in E] \}
\end{split}
\end{align}

When we draw a derivation tree for a given input $i$, we are effectively drawing the set of paths (excluding synthetic nodes) taken over the corresponding grammar's graph during expansion (hereon, $P_i \subseteq P$).
For example, the path $\nonterm{expr}_1 \rightarrow \dots \rightarrow \text{``$5$''}$ in the derivation tree shown in \cref{fig:zero-five-tree} directly corresponds to the path:
\begin{equation} \label{eq:zero-five-path}
\begin{split}
\nonterm{expr}_1 &\rightarrow |_1 \rightarrow \nonterm{number}_1 \rightarrow |_2 \rightarrow \sqcup_2 \\
                 &\hookrightarrow \nonterm{nonzero}_1 \rightarrow |_5 \rightarrow \text{``$5$''}
\end{split}
\end{equation}

$k$-path coverage maps the set of all possible inputs $I$ to the set of paths with length $k$ in the graph $H$ covered by the derivation tree of a given input, i.e. $\textbf{path}_k : I \mapsto \mathcal{P}(P)$.
In the original $k$-path definition, coverage is computed for paths of \emph{exactly} length $k$.
In practice, because some paths to terminal nodes can only exist with lengths less than $k$~\cite{kpath} (similar to $\textbf{cdrc}_n$), $k$-path is typically computed for paths of length $k$ or less, as demonstrated by the original authors~\cite{fandango,isla}; we continue with this definition.

\begin{align}
\label{eq:kpath}
\textbf{path}_k(i) &= \{ p \in P_i : \lvert p \rvert \leq k \}
\end{align}

\citet{kpath} then propose that one build a fuzzer by sampling inputs that exercise previously-uncovered paths.

\section{Empirical Results are Insufficient}\label{sec:metric-correlation}

\new{With the background now established, we motivate this work with a brief repudiation of previous evaluation practices when comparing two metrics.}
There are two versions of the $k$-path paper: one that was accepted to ASE in 2019 and an extended version published independently in 2023.
In the original paper~\cite{kpath}, the future work section suggests:
\begin{quote}
One would assume that there would be strong correlations between individual language elements and some code in the program under test...
\end{quote}
which is subsequently evaluated in the extended version~\cite{kpath-arxiv}:
\begin{quote}
... $k$-path input coverage positively correlates with code coverage [and is therefore] a useful objective for systematic generation of diverse inputs.
\end{quote}

The objective of this evaluation is to establish that as $k$-path coverage increases, so does code coverage---that is, there is a causal relationship between $k$-path and code coverage.
However, correlation is not a reliable means by which to establish a causal relationship between metrics.
To demonstrate this, we conduct a short experiment with the ``identity'' coverage metric, $\textbf{ident} : I \mapsto \mathcal{P}(I)$:
\begin{equation}
\textbf{ident}(i) = \{ i \}
\end{equation}
This is, for all intents and purposes, an input metric by the definitions from \cref{sec:metric-def}: it maps each input in an input domain to a set of coverage points (here, exactly one input).
We can now study its correlation with other coverage metrics.

\subsection{Correlating with $k$-path}\label{sec:kpath-correlation}

We experiment using the \emph{tribble} input generation tool introduced in the $k$-path paper~\cite{kpath}.
\emph{tribble} comes equipped with a ``random'' mode, allowing us to generate random inputs from a given grammar.
For direct comparisons, we use the grammars from the correlation evaluation of the extended $k$-path paper~\cite{kpath-arxiv}\footnote{We do not use the Markdown grammar, as \emph{tribble} exhausted system memory before completing a trial.}, sampling 100 inputs or until $k$-path is saturated.
We merge the data from 50 trials and perform Spearman rank correlation~\cite{spearman} between $\textbf{ident}$ and $\textbf{path}_k$.
Results of this experiment are presented in \cref{tab:kpath-correlation}.

\begin{table}
    \caption{Correlation between $\textbf{path}_k$ and $\textbf{ident}$. Bolded measurements are statistically significant ($p < 0.01$).}
    \label{tab:kpath-correlation}
    \centering
    \begin{tabular}{lrrrrr}
        \toprule
         & \multicolumn{5}{c}{Spearman's $\rho$ ($\textbf{path}_k$ vs. $\textbf{ident}$)} \\
        \cmidrule(l{0.25em}r{0.25em}){2-6}
        Grammar & $1$-path & $2$-path & $3$-path & $4$-path & $5$-path \\
        \midrule
JSON & \textbf{0.80} & \textbf{0.88} & \textbf{0.91} & \textbf{0.91} & \textbf{0.91} \\
CSV & 0.19 & \textbf{0.29} & \textbf{0.64} & \textbf{0.71} & \textbf{0.63} \\
URL & \textbf{0.89} & \textbf{0.92} & \textbf{0.93} & \textbf{0.95} & \textbf{0.96} \\
        \bottomrule
    \end{tabular}
\end{table}

Excluding CSV, we observe that the $\textbf{ident}$ coverage metric is strongly correlated with $\textbf{path}_k$.
In other words: executing more inputs increases $k$-path coverage.
This makes sense!
Since we are measuring over grammar, the probability that we cover more $k$-paths by sampling is simply an instance of the coupon collector's problem~\cite{coupon-collector}.
CSV's correlation and significance is low because random generation saturates $k$-path for small $k$ almost immediately due to the uncomplicated nature of the grammar.
This result reveals the issue with using correlation as a measure for the strength of a coverage metric: correlation reveals just as much about the sampling method as it does about the measured variables.

The covering relationships explored in \cref{sec:covering-relationships} grants greater insight into the nature of the relationships between metrics.
When we show that one metric covers another, we show that it is more \emph{sensitive}; an increase in one metric is necessarily observed as an increase in another that covers it.
This does not necessarily reveal a strong causal relationship either; after all, $\textbf{ident}$ trivially universally covers all test metrics, and we can arbitrarily construct metrics that are more sensitive than others without necessarily improving test performance.
What covering reveals that correlation cannot is that methods that efficiently saturate one metric will saturate others covered by it, and that, in doing so, we test more of each subdomain of each covered metric.

\subsection{Correlating with Code}\label{sec:correlating-with-code}

We know now that $\textbf{ident}$ correlates with $k$-path, but does it correlate with code coverage?
We reproduce Tables 2 and 5 of the extended $k$-path paper~\cite{kpath-arxiv}\footnote{Subjects jackson-databind and galimatias-nu omitted as they no longer function.} in \cref{tab:correlating-with-code} by correlating with cumulative branch coverage using the inputs from \cref{sec:kpath-correlation}.
The results are comparable with that of systematic generation, and are actually \emph{more correlated} with branch coverage than $k$-path in many cases.
Are we to take from this that $\textbf{ident}$ is a strong predictor for code coverage?
When sampling randomly, yes: this is the very basis for fuzz testing (``fuzzing'').
In the case of grammar-based fuzzing specifically, every program that parses based on a grammar contains some state machine that accepts or rejects a given input based on its conformance to that grammar.
In typical implementations, program branches that correspond to a nonterminal expansion will only be visited if its parent's branches are covered.
When we sample random derivation trees, we cover code regions corresponding to the randomly selected nonterminal expansions.
When the code is not already entirely covered, sampling more inputs will likely cover more code.

\begin{table}
    \caption{Correlation and absolute results of $\textbf{branch}$ and $\textbf{ident}$: Spearman's $\rho$ (bolded where $p < 0.01$) and arithmetic mean $\mu$ of branch coverage.}
    \label{tab:correlating-with-code}
    \centering
    \begin{tabular}{lrcc}
        \toprule
        Grammar & Subject & $\rho$ & $\mu$ \\
\midrule \multirow{11}*{JSON} & argo & \textbf{0.78} & 0.3959 \\
 & fastjson & \textbf{0.86} & 0.0360 \\
 & genson & \textbf{0.83} & 0.0880 \\
 & gson & \textbf{0.78} & 0.2272 \\
 & json-flattener & \textbf{0.62} & 0.7100 \\
 & json-java & \textbf{0.76} & 0.1629 \\
 & json-simple & \textbf{0.80} & 0.5719 \\
 & json-simple-cliftonlabs & \textbf{0.72} & 0.3729 \\
 & json2flat & \textbf{0.71} & 0.6647 \\
 & minimal-json & \textbf{0.78} & 0.4106 \\
 & pojo & \textbf{0.78} & 0.1278 \\
\midrule \multirow{6}*{CSV} & commons-csv & \textbf{0.74} & 0.3714 \\
 & jackson-dataformat-csv & \textbf{0.65} & 0.1467 \\
 & jcsv & \textbf{0.66} & 0.3622 \\
 & sfm-csv & \textbf{0.54} & 0.0721 \\
 & simplecsv & \textbf{0.69} & 0.4050 \\
 & super-csv & \textbf{0.66} & 0.1695 \\
\midrule \multirow{4}*{URL} & autolink & \textbf{0.84} & 0.5897 \\
 & galimatias & \textbf{0.87} & 0.2452 \\
 & jurl & \textbf{0.76} & 0.6678 \\
 & url-detector & \textbf{0.89} & 0.4621 \\
        \bottomrule
    \end{tabular}
\end{table}

That said, this correlation is only strong for code regions corresponding to the actual parsing.
Consider a program that then \emph{does} something with this parsed information; the code regions not pertaining to parsing would be covered merely by chance.
Relating this to the end of \cref{sec:kpath-correlation}, one would need to show that $k$-path at least subsumes the subset of code coverage corresponding to parsing in order to claim any strong relationship.
Anything beyond this is strictly a function of the distribution of inputs sampled and the probability of covering program semantics not expressed within the grammar.
Correlation does not indicate that increasing $k$ causes $k$-path to cover a subject more than simply taking more samples.
Relying on correlation of test metrics leads to conflicting results according to the sampling method used~\cite{suite-correlation,fuzz-correlation} and must not be used as justification for using one metric in place of another.

\subsection{Other Empirical Comparisons}

\new{%
\cref{sec:kpath-correlation,sec:correlating-with-code} consider correlation-based analysis to compare two metrics with constrained random input generation.
Another common method to compare metrics is to generate inputs by sampling underrepresented features of each, like in the original $k$-path work~\cite{kpath}.
While historically new metrics emerge as objectives of generation, metrics themselves are used to evaluate test suites from arbitrary sources.
Evaluations which measure the performance of generation that optimizes for some specific metric have value in empirically determining the efficacy of those generation strategies, but provide little indication as to the properties of the metrics themselves.
}

\section{Improving $k$-path Metric} \label{sec:k-alt-def}

To demonstrate how one may more rigorously compare metrics, we begin by refining $k$-path.
We show that this new metric is strictly more sensitive than $k$-path alone and requires less computational resources.

When we traverse from one node to another in the grammar graph, that means that there was a corresponding expansion step in the derivation tree.
At nonterminals and concatenations, we \emph{always} traverse the outgoing edge(s) in the graph; no matter what the next node is, it is involved in the expansion.
At alternations, however, we only traverse \emph{one} during expansion.
This means that paths other than those between alternation-variant edges are redundant, as it can be inferred from the rules of derivation.

To demonstrate the effect of this, consider the path from \cref{eq:zero-five-path}.
The \emph{information} expressed over this path by $k$-path for $k = 2$ is effectively no more than that already expressed by $k = 1$, since every node in the path is implied by either the presence of its parent or the presence of its child.
In fact, because the grammar contains no nested alternations or concatenations (and therefore no path $|_i \rightarrow |_j$ or $\sqcup_i \rightarrow |_j$), there are \emph{no inputs} for which $k$-path for $k = 2$ expresses more information than $k = 1$.
Yet, the \emph{number of paths} of length 1 is 30; the number of paths of length 2 or fewer is \emph{64}, meaning that \emph{34} of these paths are redundant.

\subsection{$k$-alt-path Definition}

The alt-paths $A \subseteq P$ of the grammar graph $H$ are the paths between and including the \emph{outgoing edges} of alternations.
That is, paths beginning with alternations and ending with immediate descendants of alternations as a result of expansion.
With $N_\text{alt} \subset N$ as the set of synthetic alternation nodes:
\begin{align}
\label{eq:altpaths}
A &= \{ x_1 \sqcup \dots \sqcup x_n \subseteq P : x_1, x_{n-1} \in N_\text{alt} \land n > 1 \} \\
\label{eq:altpaths-i}
A_i &= A \cap P_i
\end{align}

We then define $\textbf{altpath}_k : I \mapsto \mathcal{P}(A)$ like before for $k > 1$, but for paths over $k$ or fewer outgoing edges from alternations.
\begin{align}
&\textbf{countalt}(x_1 \sqcup \dots \sqcup x_n \in A) = \lvert \{ j \in [1,n) : x_j \in N_\text{alt} \} \rvert \notag \\
\label{eq:kaltpath}
&\textbf{altpath}_k(i) = \{ p \in A_i : 1 \leq \textbf{countalt}(p) \leq k \}
\end{align}

\subsection{$k$-alt-path vs. $k$-path}

Both $\textbf{path}_k$ and $\textbf{altpath}_k$ are subdomain-based criteria.
$\mathcal{SD}_{\textbf{path}_k}$ are the subdomains of $I$ corresponding to inputs for which the derivation tree traverses each path of length $k$, therefore the number of subdomains directly corresponds to the number of paths: $\lvert \mathcal{SD}_{\textbf{path}_k} \rvert = \lvert \{ p \in P : \lvert p \rvert \leq k \} \rvert$.
For $\textbf{altpath}_k$, $\lvert \mathcal{SD}_{\textbf{altpath}_k} \rvert = \lvert \{ p \in A : \textbf{countalt}(p) \leq k \} \rvert$.


\subsubsection{$k$-alt-path covers $(k+1)$-path} \label{sec:relationship}

Every path $p$ in a grammar with paths $P$ can be decomposed into subpaths, where for every subpath $p'$ and $i \in I$
\begin{equation}
\label{eq:subpath}
p' \subseteq p \land p \in P_i \Longrightarrow p' \in P_i
\end{equation}
When we subdivide the input domain by paths present in the derivation trees, the subdomain of inputs $D_p$ for which the derivation trees that contain some path $p$ are subdivided by paths that have $p$ as a prefix:
\begin{align}
Q &= \{ q : (\exists x)[p \sqcup x \in P] \} \notag \\
\label{eq:path-subdivision}
D_p &= \bigcup_{q \in Q} D_q
\end{align}
This is necessarily the case as this subdivision enumerates all paths of length $\lvert p \rvert + 1$ with $p$ as a prefix, i.e., all the cases in which the path $p$ appears.
This similarly holds over sets of paths for which $p$ is a suffix, and is recursively applicable: we can further subdivide $D_p$ by subdividing any $D_q$.

Having established that one can cover the input subdomains corresponding to the presence of paths in derivation trees, we can now show that $\textbf{altpath}_k$ covers $\textbf{path}_{k+1}$.
By the definition of paths, the logic presented at the start of \cref{sec:k-alt-def}, the expansion rules of grammars, $\forall i \in I$:
\begin{align}
\label{eq:path-implication}
\begin{split}
&x_1 \sqcup x_2 \in A_i \land x_1 \in N_\text{alt} \Longleftrightarrow \\
&\quad \quad \{ x_1 \sqcup x_2 \sqcup x'_3 \sqcup \dots \sqcup x'_n \in P \\
&\quad \quad \quad \quad : (\forall k \in [3,n))[x'_k \not\in N_\text{alt}] \} \subseteq P_i
\end{split} \\
\label{eq:start-implication}
\begin{split}
&\{ S \sqcup x'_2 \sqcup \dots \sqcup x'_n \in P \\
&\quad \quad : (\forall k \in [2,n))[x'_k \not\in N_\text{alt}] \} \subseteq P_i
\end{split}
\end{align}
$\textbf{path}_1$ and $\textbf{path}_2$ are universally covered by $\textbf{altpath}_1$.
There exists no grammar for which there exists an input $i \in I$ such that there exists some path $p \in P_i$ where $\lvert p \rvert \leq 2$ that is not implicitly present in $P_i$ due to the presence of either the start node $S$ or some alternation-child pair $x_1 \sqcup x_2 \in P_i$.
Each subdomain corresponding to each path is necessarily subdivided by some set of alt-paths, which cover each subdivision where this path is present, satisfying the covering relation by \cref{eq:path-subdivision}.
By the same logic, $\textbf{path}_{k+1}$ is universally covered by $\textbf{altpath}_k$ because any path $p \in P_i$ of length $k+1$ or less is implied by the presence of some other path $p' \in A_i$ where $\textbf{countalt}(p') \leq k$, and the subdomain corresponding to each path of length $k+1$ or less is subdivided completely.
Further, we expect that this should hold for $\textbf{path}_j$ over many $j > k+1$ because path redundancy increases as the grammar contains more nonterminal and concatenation nodes between alternation nodes.

\subsubsection{Subdivision size} \label{sec:size}

Since both $\textbf{path}_k$ and $\textbf{altpath}_k$ rely on counting unique paths, computing these metrics requires both enumeration and storage of each potential path (and therefore has storage and computation requirements proportional to $\lvert \mathcal{SD}_{\textbf{path}_k}(X,Y) \rvert$ and $\lvert \mathcal{SD}_{\textbf{altpath}_k}(X,Y) \rvert$, respectively).
For every $k$, there exists some minimum $j > k + 1$ where $\textbf{path}_j$ is \emph{not} covered by $\textbf{altpath}_k$ because $\exists p \in (P_i - A_i)$ where $\lvert p \rvert = j$ and $\textbf{countalt}(p) = k$.
By construction,
\begin{align*}
\lvert \mathcal{SD}_{\textbf{altpath}_k}(X,Y) \rvert &= \lvert \{ p \in A_i : 1 \leq \textbf{countalt}(p) \leq k \} \rvert\\
\lvert \mathcal{SD}_{\textbf{path}_{j-1}}(X,Y) \rvert &= \lvert \{ p \in P_i : 1 \leq \lvert p \rvert \leq j - 1 \} \rvert\\
\lvert \mathcal{SD}_{\textbf{altpath}_k}(X,Y) \rvert &< \lvert \mathcal{SD}_{\textbf{path}_{j-1}}(X,Y) \rvert
\end{align*}
In plain English, $\textbf{altpath}_{k}$ has lower storage requirements than $\textbf{path}_{j-1}$, the maximum $\textbf{path}$ that $\textbf{altpath}_{k}$ covers.
This directly corresponds to a reduced computational effort as the number of paths that need to be counted during a coverage computation is reduced quadratically.
Without loss of covering by \cref{eq:subpath,eq:path-subdivision}, we reduce the memory requirement even further by omitting paths contained by others.

\subsection{Evaluations on Real Grammars}\label{sec:eval}

\begin{table}
    \caption{The results of \cref{sec:eval} for increasing $k$.}
    \label{tab:results}
    \centering
    \begin{tabular}{llrrrr}
        \toprule
        Subject & $k$ & $j$ & $\lvert \mathcal{SD}_{\textbf{path}_{j-1}} \rvert$ & $\lvert \mathcal{SD}_{\textbf{altpath}_{k}} \rvert$ & Ratio \\
        \midrule
        \multirow{4}*{``Addition''}
& 1 & 4 & 110 & 17 & 6.5:1 \\
& 2 & 7 & 294 & 30 & 9.8:1 \\
& 3 & 10 & 591 & 45 & 13.1:1 \\
& 5 & 15 & 1,362 & 88 & 15.5:1 \\
        \midrule
        \multirow{4}*{CSV}
& 1 & 4 & 830 & 199 & 4.2:1 \\
& 2 & 7 & 2,360 & 298 & 7.9:1 \\
& 3 & 10 & 5,095 & 497 & 10.3:1 \\
& 5 & 16 & 20,765 & 1,661 & 12.5:1 \\
        \midrule
        \multirow{4}*{REST}
& 1 & 4 & 1,991 & 427 & 4.7:1 \\
& 2 & 6 & 4,408 & 798 & 5.5:1 \\
& 3 & 8 & 8,368 & 1,111 & 7.5:1 \\
& 5 & 14 & 41,223 & 4,662 & 8.8:1 \\
        \midrule
        \multirow{4}*{XML}
& 1 & 4 & 816 & 152 & 5.4:1 \\
& 2 & 6 & 2,105 & 350 & 6.0:1 \\
& 3 & 9 & 6,550 & 823 & 8.0:1 \\
& 5 & 15 & 31,696 & 3,411 & 9.3:1 \\
        \bottomrule
    \end{tabular}
\end{table}

To demonstrate the degree to which subdivision counts differ, we evaluate the storage requirement and covering relationship for increasing $k$ on the ``addition'' grammar of \cref{fig:addition-grammar} and the well-known context-free CSV, REST, and XML grammars\footnote{TAR is omitted as it depends on constraints to be parseable.} taken from recent works that rely on $k$-path for diversity metrics~\cite{isla,fandango}.
Our findings are presented in \cref{tab:results}.
These results emphasize the degree of reduction achieved by $k$-alt-path, which
requires 5--10$\times$ less storage for small $k$ than the corresponding $(j-1)$-path, the effect of which compounds with increasing $k$.

\subsection{Interpreting $k$-alt-path}

In \cref{sec:relationship}, we demonstrate that $k$-alt-path covers $(k+1)$-path and has fewer subdomains, therefore showing that $k$-alt-path is both more sensitive and efficient.
In \cref{sec:correlating-with-code}, however, we reasoned that the finding that $k$-path correlates with code coverage is not a strong one.
The latter would seem to undermine the former, but no: if the objective is to cover a grammar, then these metrics are perfectly suitable.
$k$-alt-path is still superior as it tests more of each subdomain of $(k+1)$-path and requires significantly less computational resources.
Nevertheless, one must not rely on grammar coverage as a proxy~\cite{silifuzz} for code metrics (e.g. branch coverage, mutant coverage~\cite{mutation-study}, or fault coverage), as correlation neither guarantees an increase in code metrics nor considers code regions that are probabilistically uncovered by sampling derivation trees from grammars.

\section{Other Input Metric Relationships}\label{sec:covering-relationships}

\begin{table}
    \caption{Explanation of covering between input metrics.}
    \label{tab:explanations}
    \centering
    \begin{threeparttable}
    \begin{tabularx}{\linewidth}{cX}
        \toprule
        Edge & Reasoning \\
        \midrule
        \phantom{0}1\tnote{$\dagger$}\phantom{\tnote{$\dagger$}} & $\textbf{uniq}(\mathcal{SD}_{\textbf{altpath}_1}) \subseteq \textbf{uniq}(\mathcal{SD}_{\textbf{rule}})$ \\
        \phantom{0}2\tnote{$\dagger$}\phantom{\tnote{$\dagger$}} & $\textbf{uniq}(\mathcal{SD}_{\textbf{rule}}) \subseteq \textbf{uniq}(\mathcal{SD}_{\textbf{path}_1})$ \\
        \phantom{0}3\phantom{\tnote{$\dagger$}}\phantom{\tnote{$\dagger$}} & See \cref{sec:alt-covers-rule}. \\
        \phantom{0}4\phantom{\tnote{$\dagger$}}\phantom{\tnote{$\dagger$}} & See \cref{sec:relationship}. \\
        \phantom{0}5\phantom{\tnote{$\dagger$}}\phantom{\tnote{$\dagger$}} & $\textbf{uniq}(\mathcal{SD}_{\textbf{path}_k}) \subseteq \textbf{uniq}(\mathcal{SD}_{\textbf{path}_{k+1}})$ (definition) \\
        \phantom{0}6\tnote{$\ddag$}\phantom{\tnote{$\dagger$}} & Shown by \citet{laemmel}. \\
        \phantom{0}7\phantom{\tnote{$\dagger$}}\phantom{\tnote{$\dagger$}} & $\textbf{uniq}(\mathcal{SD}_{\textbf{cdrc}_k}) \subseteq \textbf{uniq}(\mathcal{SD}_{\textbf{cdrc}_{k+1}})$ (definition) \\
        \phantom{0}8\phantom{\tnote{$\dagger$}}\phantom{\tnote{$\dagger$}} & $\textbf{uniq}(\mathcal{SD}_{\textbf{altpath}_k}) \subseteq \textbf{uniq}(\mathcal{SD}_{\textbf{altpath}_{k+1}})$ (definition) \\
        \phantom{0}9\phantom{\tnote{$\dagger$}}\phantom{\tnote{$\dagger$}} & See \cref{sec:alt-covers-cdrc}. \\
        10\tnote{$\dagger$}\tnote{$\ddag$} & See \cref{sec:cdrc-covers-path}. \\
        11\phantom{\tnote{$\dagger$}}\phantom{\tnote{$\dagger$}} & $\textbf{uniq}(\mathcal{SD}_{\textbf{gram}_{\leq k}}) \subseteq \textbf{uniq}(\mathcal{SD}_{\textbf{gram}_{\leq k+1}})$ (definition) \\
        12\phantom{\tnote{$\dagger$}}\phantom{\tnote{$\dagger$}} & All terminals must be covered. \\
        \bottomrule
    \end{tabularx}
    \begin{tablenotes}
        \item[$\dagger$] Requires no nested alternations or concatenations.
        \item[$\ddag$] Requires that start nonterminal has exactly one rule.
    \end{tablenotes}
    \end{threeparttable}
\end{table}

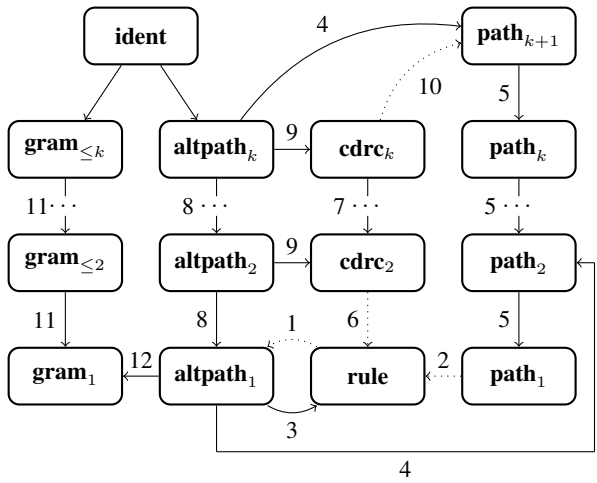
\begin{figure}
    \centering
    \small
    \begin{tikzpicture}[on grid,auto,node distance=2cm]
        \node[draw,thick,minimum width=1.5cm,minimum height=0.75cm,rounded corners]
            (rule) {$\textbf{rule}$};
        \node[draw,thick,minimum width=1.5cm,minimum height=0.75cm,rounded corners]
            (cdrc-2) [above=1.5cm of rule] {$\textbf{cdrc}_2$};
        \node (cdrc-dots) [above=0.75cm of cdrc-2] {$\dots$};
        \node (cdrc-dots-label) [left=0.375cm of cdrc-dots] {7};
        \node[draw,thick,minimum width=1.5cm,minimum height=0.75cm,rounded corners]
            (cdrc-k) [above=0.75cm of cdrc-dots] {$\textbf{cdrc}_k$};
        
        \node[draw,thick,minimum width=1.5cm,minimum height=0.75cm,rounded corners]
            (1-alt) [left=of rule] {$\textbf{altpath}_1$};
        \node[draw,thick,minimum width=1.5cm,minimum height=0.75cm,rounded corners]
            (2-alt) [above=1.5cm of 1-alt] {$\textbf{altpath}_2$};
        \node (alt-dots) [above=0.75cm of 2-alt] {$\dots$};
        \node (alt-dots-label) [left=0.375cm of alt-dots] {8};
        \node[draw,thick,minimum width=1.5cm,minimum height=0.75cm,rounded corners]
            (k-alt) [above=0.75cm of alt-dots] {$\textbf{altpath}_k$};

        \node[draw,thick,minimum width=1.5cm,minimum height=0.75cm,rounded corners]
            (1-path) [right=of rule] {$\textbf{path}_1$};
        \node[draw,thick,minimum width=1.5cm,minimum height=0.75cm,rounded corners]
            (2-path) [above=1.5cm of 1-path] {$\textbf{path}_2$};
        \node (path-dots) [above=0.75cm of 2-path] {$\dots$};
        \node (path-dots-label) [left=0.375cm of path-dots] {5};
        \node[draw,thick,minimum width=1.5cm,minimum height=0.75cm,rounded corners]
            (k-path) [above=1.5cm of 2-path] {$\textbf{path}_k$};
        \node[draw,thick,minimum width=1.5cm,minimum height=0.75cm,rounded corners]
            (k1-path) [above=1.5cm of k-path] {$\textbf{path}_{k+1}$};

        \node[draw,thick,minimum width=1.5cm,minimum height=0.75cm,rounded corners]
            (1-gram) [left=of 1-alt] {$\textbf{gram}_1$};
        \node[draw,thick,minimum width=1.5cm,minimum height=0.75cm,rounded corners]
            (2-gram) [above=1.5cm of 1-gram] {$\textbf{gram}_{\leq 2}$};
        \node (gram-dots) [above=0.75cm of 2-gram] {$\dots$};
        \node (gram-dots-label) [left=0.375cm of gram-dots] {11};
        \node[draw,thick,minimum width=1.5cm,minimum height=0.75cm,rounded corners]
            (k-gram) [above=1.5cm of 2-gram] {$\textbf{gram}_{\leq k}$};

        \node[draw,thick,minimum width=1.5cm,minimum height=0.75cm,rounded corners]
            (ident) [left=5cm of k1-path] {$\textbf{ident}$};
        
        \coordinate[below=1cm of 1-alt] (1-alt-2-path-1);
        \coordinate[right=1cm of 2-path] (1-alt-2-path-3);
        \coordinate[below=2.5cm of 1-alt-2-path-3] (1-alt-2-path-2);
           
        \path[->]
            (1-alt) edge [bend right] node [below] {3} (rule)
            (1-alt-2-path-3) edge (2-path)
            (2-path) edge node [left] {5} (1-path)
            (k1-path) edge node [left] {5} (k-path)
            (path-dots) edge (2-path)
            (cdrc-dots) edge (cdrc-2)
            (gram-dots) edge (2-gram)
            (2-alt) edge node [left] {8} (1-alt)
            (alt-dots) edge (2-alt)
            (2-alt) edge node [above] {9} (cdrc-2)
            (k-alt) edge node [above] {9} (cdrc-k)
            (k-alt) edge [bend left] node [above left] {4} (k1-path)
            (2-gram) edge node [left] {11} (1-gram)
            (ident) edge (k-gram)
            (ident) edge (k-alt)
            (1-alt) edge node [above] {12} (1-gram)
            ;
        \path[->,dotted]
            (rule) edge [bend right] node [above] {1} (1-alt)
            (1-path) edge node [above] {2} (rule)
            (cdrc-2) edge node [left] {6} (rule)
            (cdrc-k) edge [bend left] node [below right] {10} (k1-path)
            ;
        \path[-]
            (1-alt) edge (1-alt-2-path-1)
            (1-alt-2-path-1) edge node [below] {4} (1-alt-2-path-2)
            (1-alt-2-path-2) edge (1-alt-2-path-3)
            (cdrc-k) edge (cdrc-dots)
            (k-gram) edge (gram-dots)
            (k-path) edge (path-dots)
            (k-alt) edge (alt-dots)
            ;
    \end{tikzpicture}
    \caption{Covering relationships between input metrics. $C_1$ covers $C_2$ if there is a path from $C_1$ to $C_2$. Restrictions from \cref{tab:explanations} denoted with dotted edges.}
    \label{fig:relationships}
\end{figure}

Now that we have motivated use of the covers relationship, we now review the relationships between all input metrics discussed in this paper.
These are enumerated in \cref{tab:explanations} and \cref{fig:relationships}.
Most of these relationships may be shown simply by comparing their subdomains; if $\mathcal{SD}_{C_2}$ is a subset of $\mathcal{SD}_{C_1}$, there is some subset of the subdomains of $C_1$ that covers the subdomains of $C_2$.
For example: $\textbf{rule}$ covers $\textbf{altpath}_1$ because every rule is either a variant of an alternation, or is the only rule for that nonterminal, at least when nested alternations are forbidden.
$\textbf{path}_1$ covers $\textbf{rule}$ for similar reasons; by the construction steps of the grammar graph in \cref{sec:graph-construction}, there is at least one node that corresponds exactly to each expansion of each nonterminal.
The remaining trivial covering relationships follow by their definition.

\subsection{Non-trivial Covering Relationships}

Not every covering relationship can be shown by comparing subdomains directly.
To show that a metric $C_1$ covers $C_2$, we must show that, for each subdomain of $C_2$, there exists some subset of subdomains of $C_1$ that, when unioned, are \emph{exactly} that subdomain.
We explain why this holds for several pairs of metrics in the following subsections, though with less detail as they are each consequences of \cref{sec:relationship}.

\subsubsection{1-alt-path covers Rule Coverage} \label{sec:alt-covers-rule}

Each nonterminal in a well-formed grammar has one to many associated rules.
When there are more than one associated rules, we represent this with an alternation; as a result, there is exactly one edge in the grammar graph corresponding to each variant in this alternation, and $\textbf{altpath}_1$ and $\textbf{rule}$ share a corresponding subdomain of inputs formed by the paths to this edge.
When a nonterminal has only one associated rule, the subdomain corresponding to the expansion of this rule is subdivided by the previous rule expansions or the start node; this is a result of \cref{eq:path-implication,eq:start-implication}.
Therefore, $\textbf{rule}$ is covered by $\textbf{altpath}_1$, because there is no rule for which there is not some subdividing set of alternation-variant pairs.

\subsubsection{$k$-alt-path covers CDRC-$k$} \label{sec:alt-covers-cdrc}

$\textbf{cdrc}_k$ is effectively the chained version of rule coverage, like $\textbf{altpath}_k$ is the extended form of alternation coverage.
Every sequence of expansions is the union of subdomains formed by subsets of sequences of alternation-variant pairs in the grammar graph (i.e., the rule expansions), just like $\textbf{rule}$.

\subsubsection{CDRC-$k$ covers $(k+1)$-path} \label{sec:cdrc-covers-path}

Expansion sequences correspond quite well to paths in the grammar graph when forbidding nested alternations and concatenations.
An expansion corresponds to two entries in the graph: the nonterminal and its immediate descendant (when the nonterminal has exactly one rule) or the nonterminal's alternation and its immediate descendant.
When considered together, a sequence of $k$ expansions corresponds to, minimally, paths of length $k+1$ in the grammar graph; for the same reasons as the previous two metric relationships and \cref{sec:relationship}, this suggests that $\textbf{cdrc}_k$ covers $\textbf{path}_{k+1}$, but not vice versa, as the introduction of any alternations or concatenations increases this path length.



\subsection{Lexical vs. Syntactic Coverage}\label{sec:lexical-vs-syntactic}

\cref{tab:explanations} and \cref{fig:relationships} reveals no relationships between $\textbf{gram}_{\leq n}$ and any syntactic input metrics for $n \geq 2$.
The reason for this is simple: there are no universal covering relationships between $\textbf{gram}_{\leq n}$ and any syntactic metric for any $n \geq 2$.
We may demonstrate this by counterexample.

\subsubsection{$n$-or-less-gram does not subsume rule coverage}

Recall that subsumption is implied by covering; if a metric cannot subsume another, then it also cannot cover it.
$\textbf{rule}$ is the weakest syntactic coverage; if we cannot subsume $\textbf{rule}$, then we cannot subsume any other syntactic metric.
Choose an arbitrary $n \geq 1$; we may construct a grammar $\langle \{S\}, \Sigma, \{S\} \times \Sigma^{n+1}, S \rangle$.
For $\textbf{gram}_{\leq n}$ to subsume $\textbf{rule}$, its satisfaction must imply the satisfaction of $\textbf{rule}$, but it trivially does not.
We may select an arbitrary terminal symbol $x \in \Sigma$ and a subset of inputs $I' = \{ x \sqcup y : y \in \Sigma^n \} \subset \Sigma^{n+1} = I$.
$I'$ saturates $\textbf{gram}_{\leq n}$, but $I$ is the minimum set of inputs to saturate $\textbf{rule}$; since $I' \subset I$, $\textbf{gram}_{\leq n}$ does not subsume $\textbf{rule}$.

\subsubsection{$k$-alt-path does not subsume 2-or-less-gram}

$\textbf{altpath}_k$ is the strongest syntactic metric defined in this paper; if it cannot universally cover $\textbf{gram}_{\leq 2}$, no syntactic metric defined in this paper can.
For arbitrary $k$, we can construct a grammar $G = \langle V, \Sigma, R, S \rangle$ such that:
\begin{align*}
V &= \{ u, v_1, \dots, v_k \} \\
\Sigma &= \{ x_1, x_2 \} \cup \{ y_1, \dots, y_k \} \\
R &= \{ (S, uv_1), (u, x_1), (u, x_2) \} \\
&\cup \{ (v_j, y_j) : j \in [1,k] \} \cup \{ (v_j, v_{j+1}) : j \in [1,k) \}
\end{align*}

As a result, $I = \{ x_1, x_2 \} \times \{ y_1, \dots, y_k \} \subset \Sigma^2$; only $I$ will saturate $\textbf{gram}_{\leq 2}$.
$I' = \{ x_1y_1 \} \cup \{ x_2y_j : j \in [2,k] \}$ saturates $\textbf{altpath}_k$.
Since $I' \subset I$, $\textbf{altpath}_k$ does not subsume $\textbf{gram}_{\leq 2}$.



\section{Discussion}

Before concluding, we discuss the limitations, threats to validity, the implications of this paper, and related work that covers similar ground to ourselves.

\subsection{Limitations}

$k$-alt-path, like $k$-path, is not well-defined for CFG extensions like repetitions.
Furthermore, while we establish the covering relationship for input metrics, covering alone does not indicate superiority in fault detection.
Those familiar with the work of \citet{subsumption} will notice that this paper does not attempt to establish the \emph{properly covers} relationship between metrics.
A given test adequacy metric properly covers another if it covers \emph{uniquely}, where no subdomain is used more than once in the covering of the subdomains of the other metric.
$k$-alt-path trivially does not properly cover $k$-path simply because $k$-path has more subdomains.
This distinction is crucial, as one can establish that if one metric properly covers another, then it is also necessarily superior at fault detection~\cite{subsumption} under the failure rate model~\cite{measure}, whereas this is not the case for the covering alone.
That said, the proof for this claim is in the context of partition testing, where each subdomain is sampled exactly once regardless of whether that subdomain has already been covered by other samples.
This is not the model under which we test programs with random inputs, where we either (1) generate inputs without regard to input subdomains, or (2) continuously produce inputs.
We suggest that covering likely \emph{does} indicate superior fault detection under the failure rate model in fuzzing, but proving this would require a better model of fuzzer behavior and may fail to hold for certain classes of fuzzers (e.g., mutational coverage-guided fuzzers~\cite{fuzzing-survey}).
As such, we defer this to future work.

\subsection{Threats to Validity}

\cref{tab:results,tab:kpath-correlation,tab:correlating-with-code} are results from empirical evaluations.
As such, they must be caveated with potential threats that may affect their correctness and generality.

\subsubsection*{External}
The selection of grammars in \cref{sec:eval} was based on grammars present in other evaluations of works utilizing the $k$-path.
We do not claim that these results are representative for all grammars; indeed, we can construct grammars for which the resource requirement ratios between $k$-alt-path and $k$-path are significantly lower.
The general, weaker claims of minimal $j$-path not covered by $k$-alt-path and resource requirements are justified in \cref{sec:relationship} and \cref{sec:size}, respectively.

Similarly, the correlation of the identity metric with $k$-path and branch coverage presented in \cref{sec:kpath-correlation,sec:correlating-with-code} may only be the case for these grammars, which were selected due to their presence in a previous work~\cite{kpath-arxiv}.
Only showing that this correlation is present for a few grammars serves as a counterexample to the claim that strong correlation indicates a strong causal relationship between metrics, obviating this threat.

\subsubsection*{Internal}
The findings of \cref{sec:kpath-correlation,sec:correlating-with-code} are subject to the random set of inputs produced by \emph{tribble}~\cite{kpath}.
In degenerative cases, the correlation might be weaker than presented here.
We control for this by performing several trials, and reiterate that these measurements serve as a counterexample rather than evidence for a general claim.

\subsubsection*{Construct}
The findings of \cref{sec:kpath-correlation,sec:correlating-with-code} use the random generator of \emph{tribble} configured with a maximum depth of 100.
This implements a depth-limited derivation tree generator with random decisions local to alternations being sampled.
Other methods for random derivation tree generation exist, so while this is the most typical and straightforward implementation, these results may not represent all forms of random derivation tree sampling.
Additionally, we sample up to 100 inputs in both experiments.
The choice of the number of samples is arbitrary, and other choices may reveal differing strengths of correlation.
Given that this experiment serves primarily as an example of how correlation can be misleading, we believe that neither of these threats invalidate our findings.

\subsection{Implications for Modern Fuzzing}

In \cref{sec:eval}, we consider grammars from works that prototype a relatively new class of fuzzing, called \emph{language-based testing} (LBT)~\cite{isla,fandango}.
These fuzzers use grammars in combination with constraints to produce inputs that satisfy requirements beyond syntactic validity.
An open question in LBT \new{and other modern input generation strategies} is how one produces \emph{diverse} inputs; while both \textsc{ISLa}~\cite{isla} and \textsc{Fandango}~\cite{fandango} achieve some amount of syntactic diversity as measured by $k$-path, it is so far unclear whether this (in the context of their constrained generation) corresponds to an increased semantic diversity.
Indeed, for \textsc{Fandango} in particular (which uses evolutionary search to satisfy constraints), this question is critical for ensuring satisfaction of multitudinous constraint systems~\cite{niching}.
In the same way that lexical and syntactic diversity have no covering relationships as explored in \cref{sec:lexical-vs-syntactic}, the same is likely true for semantic diversity (e.g., there are likely many \new{desirable metrics that would not be captured by grammar-agnostic ones}).
\new{The misalignment of generation objectives and the metrics used to diversify inputs in generation processes of LBT, LLM-based test generation, and other specialized generators demands future work that develops metrics that can truly measure the \emph{semantic} diversity of test cases}.

\new{Furthermore, our work demonstrates that the comparison of metrics must be heavily scrutinized.
Correlation of metrics spontaneously emerging merely as a result of sampling indicates that conclusions based on correlation must be extremely specific as to the sampling pattern used and denote that results are not general.
Comparisons that consider performance (e.g., by Mann-Whitney U, as per fuzzing evaluation standard~\cite{prudent-practices}) must ensure to compare against reasonable baselines (e.g. \textbf{ident} presented here) along comparable objectives (e.g., time or number of generated inputs), but ultimately compare generation techniques, not metrics.
Works performing metric or technique comparisons should also consider set-theoretic analyses that do not remove nuance about how these metrics are covered~\cite{sbft}, as aggregate comparisons can obscure the differences between approaches.}

\subsection{Related Work}

\subsubsection*{Partition Testing}
Our work retreads numerous findings; \cref{sec:metric-correlation} mostly discusses known consequences of comparing metrics with statistical modeling~\cite{test-size,gen-dangers}.
The network of covering relationships presented in \cref{sec:covering-relationships} was inspired largely by \citet{subsume-relations}, which builds a network for the subsume relation between various code coverage metrics.
Presenting new metrics that improve upon existing metrics and proving their relationships is a motif of partition testing literature~\cite{subsume-survey}, and we happily extend this tradition to input metrics.
We hope this paper will act as a bridge for new research in input generation techniques to engage with partition testing works, which provide methods to reason about new metrics.

\subsubsection*{Statistical fuzzer modeling}
Likewise, we believe our findings are related to other recent studies of test adequacy metrics in fuzzing literature.
\citet{fuzz-correlation} showed recently that, while code coverage and fault coverage are \emph{very strongly correlated}, there is \emph{no strong agreement} between these metrics.
In other words, while one can predict that a fuzzer with high coverage will also find more faults, one cannot predict the ranking of a collection of fuzzers by fault coverage from the ranking by code coverage.
While initially puzzling, this result is, in essence, the same observation of our analysis in \cref{sec:correlating-with-code}: while increasing aggregate code coverage guarantees that more code will be tested, it says nothing about \emph{which} code regions are tested nor the \emph{depth} to which those code regions are tested.
Moreover, the correlative result shares further similarity with \cref{sec:kpath-correlation}: while code coverage correlates with fault coverage, this also says nothing about the degree to which code coverage covers fault coverage.
In the context of test suites, where there is greater dissimilarity between individual suites, this correlation weakens~\cite{suite-correlation}.

\subsubsection*{Other input metrics}
Though we discuss several syntactic input metrics in this paper, there are a few niche metrics which are not compared in this paper due to dissimilarity in their design.
There are quite a few such works; we review the most prominent or related to this work below.

From the syntactic approach, \citet{combinatorial} considers combinatorial coverage, or the coverage of entire derivation trees up to a certain depth or at certain boundaries.
This trivially covers many of the path-based metrics presented in this paper, but is exponential with depth.
To refine this, \citeauthor{combinatorial} also consider a number of ``control mechanisms'' which mitigate this exponentiality at the cost of some completeness.
\citet{io-grammar} defines a grammar coverage metric based on the pairs of grammar elements encountered during pre-order traversal of the derivation tree, giving a ``horizontal'' coverage compared to the ``vertical'' syntactic metrics discussed in this paper.
It additionally defines synthetic nonterminals to represent certain fields present in binary context-sensitive grammars, whereas the grammars presented in this paper are restricted or reduced to context-free.

Semantic coverage metrics have been defined, but are often too complex to see widespread adoption.
\citet{cov-comptest} introduce a number of semantic coverage metrics specific to compiler testing, namely those based on abstract state machines (historically, ``evolving algebras'')~\cite{asm} and tree-finite state machines~\cite{montages}.
This work then generates test cases which increasingly satisfy these metrics, finding several faults.
As far as we could find, there are no works which subsequently attempt to use these metrics.
\citet{spectest} identified (though do not experiment to confirm) that the computational cost of these metrics were prohibitive, and improve upon them by making their measurement tractable for large-scale testing. 
They do not relate the strength of their metric against that of \citet{cov-comptest}, but their prototype reveals a number of faults in the Java and Solidity compilers with relatively few test cases, taking several weeks to generate and execute.
So far, we could not find any work that has evaluated whether guidance by input semantic coverage metrics meaningfully improve fault coverage over syntactic coverage metrics or traditional mutational code coverage-guided fuzzers alone.
These questions remain open for future works which generate highly constrained inputs.


%

\section{Conclusion}

%

\new{%
In this paper, we outline, justify, utilize, and systematize existing partition testing techniques for comparing input-based test adequacy criteria (i.e., input metrics).
We do this to create a foothold for future research in developing input metrics for new problems facing the testing field, like those of complex input generation.
In the course of our work, we significantly improve $k$-path with $k$-alt-path, doing so using partition testing analysis to identify desirable metric properties, namely concision and sensitivity.
For now, we hope that this work serves to bring interest to the development of new metrics, identifying and remediating weaknesses in those that already exist and exploring new ways to describe and understand our test suites.
In future work, we plan to use this analysis further by developing language-specific semantic metrics and integrating them into modern input generation tools.
}

\section*{Artifact Availability}

All code used in this work may be found on Zenodo:
\begin{center}
    \url{https://doi.org/10.5281/zenodo.19251724}
\end{center}
This artifact directly constructs \cref{tab:results,tab:kpath-correlation,tab:correlating-with-code} and includes experimental data which was generated or measured during our evaluation for reference.
This code should execute on any modern consumer Linux computer and is not affected by performance.

\ifanonymous
\else
\section*{Acknowledgments}

A deep thank you to all my colleagues who reviewed this work, especially Keno Hassler, Jasper von der Heidt, Moritz Schloegel, Sahil Sihag, Alexi Turcotte, Xinyi Xu, Jos\'e Antonio Zamudio Amaya, and Andreas Zeller, who suffered through far denser, rougher drafts.
A second thanks to Andrew Fryer, Benjamin Kushigian, and Marcel B\"ohme for their insightful comments which eventually led to the creation of this paper.
I would also like to thank my supervisor, Thorsten Holz, and the Saarbrücken Graduate School of Computer Science for their respective support in my doctoral studies.
\fi

\ifanonymous
\clearpage
\fi

\bibliographystyle{IEEEtranN}
\bibliography{references}

\end{document}